\documentstyle[psfig, epsfig, 12pt]{article} 
\input{epsf.sty}

\def\mypagenumber{1}

\def\myend{\end{document}}

\normalsize

\newcounter{sxn}

\newcounter{axn}

\date{}

\newdimen\mybaselineskip
\newcommand{\beeq}{\begin{equation}}
\newcommand{\eneq}{\end{equation}}
\newcommand{\be}{\begin{eqnarray}}
\newcommand{\ee}{\end{eqnarray}}
\newcommand{\bpic}{\begin{picture}}
\newcommand{\epic}{\end{picture}}

\def\d{\partial}

\def\la{\raise.16ex\hbox{$\langle$} \, }
\def\ra{\, \raise.16ex\hbox{$\rangle$} }

\def\psibar{ \psi \kern-.65em\raise.6em\hbox{$-$} }
\def\mbar{ m \kern-.78em\raise.4em\hbox{$-$}\lower.4em\hbox{} }

\def\L{ {\cal L} }

\def\n@space{\nulldelimiterspace=0pt \mathsurround=0pt }
\def\huge#1{{\hbox{$\left#1\vbox to 20.5pt{}\right.\n@space$}}}

\def\myskip{\noalign{\kern 8pt}}
\def\myeqspace{\noalign{\kern 10pt}}

\def\boxit#1{$\vcenter{\hrule\hbox{\vrule\kern3pt
    \vbox{\kern3pt\hbox{#1}\kern3pt}\kern3pt\vrule}\hrule}$}
\def\bigbox#1{$\vcenter{\hrule\hbox{\vrule\kern5pt
     \vbox{\kern5pt\hbox{#1}\kern5pt}\kern5pt\vrule}\hrule}$}

\def\ignore#1{{}}

\begin{document}

\bibliographystyle{unsrt}
\footskip 1.0cm

\thispagestyle{empty}
\setcounter{page}{\mypagenumber}

             
\begin{flushright}{UFIFT-HEP-01-4\\}
\end{flushright}
\begin{flushright}{OUTP-01-26-P\\}
\end{flushright}

\vspace{2.5cm}
\begin{center}
{\LARGE \bf {Phase Transition in Compact QED$_{3}$ and the Josephson Junction  
}}\\ 

\vspace{2cm}
{\large Vakif K. Onemli$^{a,}$\footnote{e-mail:~ 
onemli@phys.ufl.edu},
\hskip 0.3 cm 
 Murat Tas$^{b,}$\footnote{e-mail:~ 
tas@metu.edu.tr},
\hskip 0.3 cm 
Bayram Tekin$^{c,}$\footnote{e-mail:~
tekin@thphys.ox.ac.uk}\\
\vspace{.5cm}}
$^a${\it Physics Department, University of Florida, Gainesville, 
FL 32611, USA} \\  
\vspace{.5cm}
$^b${\it Physics Department, Middle East Technical University, Ankara, 
Turkey}\\  
\vspace{.5cm}
$^c${\it Theoretical Physics, University of Oxford, 1 Keble Road, Oxford,
OX1 3NP, UK}
\end{center}

\vspace*{2.5cm}


\begin{abstract}
\baselineskip=18pt
We study the finite temperature phase transition 
in  2+1 dimensional compact QED and its dual theory: Josephson junction.
Duality of these theories at zero temperature was established 
long time ago in \cite{hosotani}. 
Phase transition in compact QED is well studied thus we employ the
`duality' to study the superconductivity phase transition in
a Josephson junction. For a thick junction we obtain a critical
temperature in terms of the geometrical properties of the junction.
\end{abstract}
\vfill
  
 
\newpage



\normalsize
\baselineskip=22pt plus 1pt minus 1pt
\parindent=25pt

\section{Introduction}

Through Polyakov's seminal works \cite{polyakov} in 2+1 dimensional compact QED and the
spontaneously broken Georgi-Glashow model; we have learned that once the 
effects of non-perturbative objects (monopole-instantons) are taken 
into account gauge theory vacuum behaves like a dual superconductor \cite{mandelstram}
which confines electric charges. Long range order in the vacuum is destroyed
by the condensation of instantons (which look like the four dimensional monopoles).
Even though 2+1 dimensional model is too simple to describe the `confinement' problem
of realistic QCD, the underlying physics in Polyakov's theory is extremely rich and
potentially useful for four dimensional physics. To give one example, it was
proposed in \cite{kogan} that chiral phase transition in QCD resembles to the deconfining
phase transition in Polyakov's model. In this paper we shall make an other use of this model.

After Polyakov's work, Hosotani \cite{hosotani} 
wrote an interesting paper not only demonstrating the `dual superconductor' picture of the
2+1 dimensional gauge theory vacuum but also refining the notion of a dual superconductor
in this context.
Namely he showed that compact QED vacuum
is dual to the barrier region in a Josephson Junction (JJ) instead of a `usual' one-piece
superconductor. In a JJ 
a normal barrier (non-superconducting) placed between two superconductors 
{\it{becomes}}
superconducting in response to the supercurrents  that flow
through the barrier \cite{josephson}. If one inserts a monopole and an anti-monopole pair
into the barrier one should observe a linear potential between the pair instead of a 
logarithmic one. A magnetic flux tube is formed and the barrier region confines the monopoles. 
One can formulate a duality
between the compact QED and the barrier region of a (dual) Josephson Junction.
Following Hosotani, we will sketch the details of this duality below but for now we should 
mention that supercurrents in the JJ correspond to the instantons in compact QED.

Our aim in this paper is to study the finite temperature phase transition
of JJ through the above mentioned duality. Strictly speaking we shall be
interested only with superconductor-normal metal-superconductor (SNS) junctions 
with {\it{thick}} metal barriers instead of superconductor-insulator-superconductor 
(SIS) junctions which cannot be made so thick.  The computations in compact QED are valid
in weak coupling and we shall see that its dual theory (SNS) junction should have
quite a thick ($\sim 100 \mu m$) barrier.

In the context of finite temperature phase transition
it is of extreme importance to make a distinction between compact QED and the
spontaneously broken ($SU(2) \rightarrow U(1)$) Georgi-Glashow model even though
these theories look the same at zero temperature.~\footnote{This distinction was not
observed in \cite{hosotani} since at zero temperature W-bosons are bound in pairs 
and do not effect the low energy dynamics.} The latter theory
accommodates charged (dynamical) particles (W-bosons) whereas the former does not.
Since there are no dynamical monopoles in JJ (monopoles in the barrier are put by hand)
; it can not be dual to Georgi-Glashow 
model. Hence JJ can be dual to compact QED which has a gauge field defined on a compact 
interval. As in Montonen-Olive duality \cite{montonen} electric charges in the JJ are dual to the 
magnetic charges in compact QED
which arise as topological objects. But bearing in mind that in JJ electric charge
is dimensionless whereas in 2+1 dimensional compact QED magnetic charge 
(inverse of the gauge coupling) is dimensionful we will need to clarify the 
previous statement.

As an example of qualitative and quantitative differences between the Georgi-Glashow model
and compact QED it was demonstrated in \cite{dunne} that the deconfining phase transition
in the former model is in the universality class of the Ising model whereas
deconfining in the compact QED is that of  Berezinskii-Kosterlitz-Thouless type
\cite{agasiyan,dunne} and the actual critical temperatures of these phase transitions
are different.

\section{Compact QED $\sim$ Josephson Junction}

Compact QED defined by the (Euclidean) path integral
\be
{\cal{Z}} = \int {\cal{D}}\, A_\mu\, {\mbox{exp}}\{-{1\over 4g^2}\int d^3x F_{\mu \nu}F^{\mu
\nu}\}  
\label{model1}
\ee
with $O(2)$ gauge symmetry has a low energy description in the weak coupling
in terms of a massive scalar field $\chi$  with the following effective 
partition function~\footnote{A proper formulation of the theory can be carried out
on the lattice but here we do not wish to dwell on this } 
\beeq
{\cal{Z}}_{eff}=\int {\cal{D}}\,\chi\, {\mbox{exp}}\{-{g^2\over 32\pi ^2}\int
d^3x  \{(\partial_\mu\chi)^2 + M^2 \cos \chi \}\}
\eneq
where $M$ is the dynamically generated $\chi$-field mass due to the Debye screening of monopole-instantons and parametrically it is related to monopole fugacity. Therefore it is 
much much smaller than $g^2$.

The electromagnetic field of the monopole-instantons
defined as $H_\mu (x)\equiv
\frac{1}{2}\epsilon_{\mu\nu\sigma}F_{\nu\sigma}$ is
computed to be~\footnote{Our normalization of $H_\mu$
is different from the normalization of \cite{hosotani}. We keep the dimension of
the electric and magnetic fields to be $mass^2$ 
both in the compact QED and Josephson Junction}
\be
H_\mu=i\frac{g^2}{4\pi}\partial_{\mu}\chi (x)
\ee
One can analytically continue the fields to Minkowski space as $H_\mu
=(F_{23},F_{31},F_{12})=i(-E_2,E_1,-iH)$ which yields
\beeq
\{H,E_1,E_2\}=\frac{g^2}{4\pi} \{{\partial \over \partial t},- {\partial \over \partial x_2},
{\partial \over \partial x_1} \} \chi
\eneq

Together with the sine-Gordon equation these are the equations of compact QED and next we turn our attention to
the Josephson Junction. When a Josephson junction is connected to a DC source a pair current
density $J$ is driven through the barrier:
\beeq
J=J_{c}\sin {\phi}
\eneq
where $J_{c}$ is the maximum supercurrent density
that the junction can support and $\phi$ is the phase difference of
Landau-Ginzburg wave function
in the two superconductors.
The astonishing feature of the Josephson junctions occurs in the presence
of zero voltage difference. Quantum mechanical nature of the phenomenon
provides us a DC current via a constant (not necessarily zero) phase
difference. Therefore $\phi$, the phase difference of the pair wave
function between the two superconductors is the key parameter of the
Josephson effect. Assuming that the $x^3$ direction is parallel to the
normal of the barrier one obtains the following equations in the 
junction \cite{josephson} 
\beeq
\{E_3,H_1,H_2 \}={1\over 2e( \lambda_1 +\lambda_2 + d)} \{{\partial \over \partial t},- {\partial \over \partial x_2},
{\partial \over \partial x_1} \} \phi
\eneq
where $d$ is the thickness of the barrier and $\lambda_1$ and  $\lambda_2$ are the penetration depths of superconductors. Maxwell's equation for JJ yield the
sine-Gordon
\beeq
(\d ^2_{x}+\d ^2_{y}-\frac{1}{v^2}\d ^2_{t})\phi
=\frac{1}{\Lambda^2_J}\sin \phi
\eneq
where $v$ and Josephson penetration depth $\Lambda_{J}$ are given in terms of the 
properties of the junction and the fundamental constants.

It is evident that both compact QED and JJ are described by
similar sets of equations. It was proposed in \cite{hosotani} that 
these theories are dual given that electric fields in compact QED are
replaced by magnetic fields in the Josephson junction. As usual this electric-magnetic
duality works only if one exchanges electric-magnetic charges \cite{montonen}. But one should
realize that the electric charge in the Josephson junction is dimensionless
but on the other side magnetic charge, which is proportional to $1/g$, in the 3D compact QED
is dimensionful. Therefore a naive correspondence between $1/g$ and $e$ is not possible.
A proper way to formulate the correspondence of these two theories is to
conjecture the following identification
\be
\chi = \phi  \hskip 1 cm \mbox{and} \hskip 0.5 cm \{H,E_1,E_2\} =\{E_3,H_1,H_2 \}
\ee
This identification leads to
\be
\frac{g^2}{4\pi} = {1\over 2e( \lambda_1 +\lambda_2 + d)}
\label{duality}
\ee
We therefore assume that given the above relations compact QED
{\it{describes}} Josephson Junction. This identification clearly is
stronger than the duality of these two theories.

\section{Phase Transition In Josephson Junction}
One can 
assume that the correspondence we have suggested above, which is valid at zero temperature,
continues to hold at finite temperature and study the phase transitions in these theories.
In compact QED one expects that at a certain temperature instantons are bound in pairs and their
effect in the partition function is suppressed. At this point the 
deconfining sets in and the gauge theory vacuum is ordered. 
In Josephson Junction above a certain temperature the supercurrents cease
to exist and the barrier region looses its `superconductivity'. The deconfining phase transition
in compact QED is better understood \cite{agasiyan,dunne} and in what follows we will study
the phase transition in JJ through the deconfining phase transition of the compact QED.
Let us recap briefly what happens in compact QED  at high temperature.

At zero temperature monopoles interact with a three dimensional
Coulomb interaction but at finite temperature 
interaction becomes logarithmic at distances 
larger than the inverse temperature $(1/T)$. This follows
from the fact that the path integral is formulated
with periodic boundary conditions in the Euclidean time
direction which becomes compact at finite temperature.
Therefore the magnetic field of an instanton 
is effectively squeezed to two dimensions when looked from
far away. The instanton density, which is proportional to the 
photon mass is so small that the average distance between the
instantons is much much bigger than $1/T$. Therefore one can
dimensionally reduce the theory and obtain a 2D
sine-Gordon theory
\be
\L = {g^2\over32 \pi^2 T}(\d_{i}\chi)^2 + {M^2 g^2\over 16\pi^2 T}\cos
\chi.
\label{sinG}
\ee
This Lagrangian describes a two dimensional Coulomb gas and it has been extensively studied 
as an exactly solvable theory. In particular it is well known that it  
undergoes a Berezinskii-Kosterlitz-Thouless \cite{bkt} phase transition.
At a temperature
\be
T_{BKT}={g^2\over 2\pi}
\label{TBKT}
\ee
monopole-anti-monopole pairs bind to form `molecules'. 
The conformal dimension of the cosine term is
\be
\Delta = {4\pi T\over g^2}
\ee
Therefore above $T_{BKT}$ the interaction term is irrelevant and the (dual)
photon becomes massless. Even though we have given a rather cursory
account of the story, detailed study \cite{agasiyan} shows that
deconfining phase transition in compact QED is that of BKT type.

In the Josephson Junction, according to the duality arguments of Hosotani \cite{hosotani},
we expect a similar phase transition at $T_{BKT}$. Namely above this temperature
supercurrents in the barrier are not freely flowing or proximity effects
are suppressed. In terms of the properties of the Josephson Junction one can
compute this temperature making use
of the duality equation (\ref{duality}). 
Therefore we obtain a phase transition temperature for the Josephson Junction
(at zero external magnetic field).
\be
T_{JJ}= {1\over e( \lambda_1 +\lambda_2 + d)}
\ee 
This formula can only be valid for clean and sufficiently thick junctions.
Taking $d = 100 \mu m$ and neglecting $\lambda_{1,2}$ one obtains
$T_{JJ}\sim 76 K$. Although for high temperature superconductors this 
temperature is not unreasonable,
it is still two orders of magnitude larger than what one obtains 
from experiments \cite{hsiang}.  
This discrepancy is not surprising given the
simplicity of our approach. We have assumed a perfectly clean, infinitely wide metal and
neglected all the complicated physics of finite temperature effects. 
Strictly speaking one should read our formula as giving an upper limit of
the transition temperature. 

\section{Conclusion}
We have studied the phase transitions in 
Compact QED and the Josephson Junction by making use of the electric-magnetic
duality suggested in \cite{hosotani}. Since both theories are described by
2D Coulomb gas at finite temperature the phase transitions are that 
of BKT type and one can obtain the critical temperatures. 
Even though our computation is quite simple, having neglected many subtle issues
that arise in finite temperature superconductors, we think that compact QED
broadly describes the physics of large S-N-S junctions.

Finally as we have alluded to it above: two dimensional sine-Gordon theory
defined by Eqn. (\ref{sinG}) is exactly solvable [See \cite{zamo} and references therein].
In particular one has the full knowledge of the soliton and breather solutions
of the theory (with Minkowski signature) as well as various correlation functions of the
theory. One is tempted to make use of this field theory knowledge to understand 
the physics of the Josephson Junction. In fact there is a vast amount of theoretical and
experimental work on solitons ( fluxons ) \cite{barone, miller} which
discuss the emergence of solitons in the context of JJ. In this work we have 
have refrained from discussing solitons since we are considering finite 
temperature theory where everything is de facto time independent (in fact 
we are in the Euclidean theory).
There are two dimensional `instanton' (rather than soliton) solutions to 
the finite temperature theory, Eqn. (\ref{sinG}). But these instantons have 
infinite action and so they are suppressed in the quantum theory at temperatures
until entropy dominates over the action. Near the phase transition temperature these
instantons play their role and one should see their effects in the physical observables
which depend on the field $\chi$ such as the supercurrent etc. Although these
issues are worth discussing from a theoretical point of view, in the actual experiments
junction irregularities and various losses are quite important to the extend that
their effects render the simple sine-Gordon picture insufficient as we have seen 
from the large value of the predicted phase transition temperature. A proper
description should take the losses into account.

\section{Acknowledgments}

We would like to thank J. H. Miller for informing us about the experiments 
on solitons in Josephson Junctions.
V.\ K.\ O. is
supported by DOE Grant DE-FG02-97ER41029.
B.\ T.\ is supported by  PPARC Grant PPA/G/O/1998/00567.




\myend